\documentstyle[12pt,fleqn,espcrc1,epsfig]{article}

\begin{document}

\noindent
{\bf {Multiplicity Distributions and Charged-neutral Fluctuations}}

\bigskip
\bigskip

\noindent
{\normalsize{Tapan K. Nayak$^c$} 
\footnote{email : nayak@veccal.ernet.in} ~~~~~~~ \\
\it (for the WA98 Collaboration) }

\medskip
\noindent
{\small M.M.~Aggarwal$^{a}$, A.~Agnihotri$^{b}$, Z.~Ahammed$^{c}$,
A.L.S.~Angelis$^{d}$, V.~Antonenko$^{e}$, 
V.~Arefiev$^{f}$, V.~Astakhov$^{f}$,
V.~Avdeitchikov$^{f}$, T.C.~Awes$^{g}$, P.V.K.S.~Baba$^{h}$
% ...
}

\smallskip
\noindent
\small\it{$^{a}$Univ. of Panjab (India)}
\small\it{$^{b}$Univ. of Rajasthan (India)}
\small\it{$^{c}$VECC, Calcutta (India)}
\small\it{$^{d}$Univ. of Geneva (Switzerland)}
\small\it{$^{e}$Kurchatov (Russia)}
\small\it{$^{f}$JINR, Dubna (Russia)}
\small\it{$^{g}$ORNL, Oak Ridge (USA)}
\small\it{$^{h}$Univ. of Jammu (India)}
\small\it{$^{i}$Univ. of M{\"u}nster (Germany)}
\small\it{$^{j}$SUBATECH, Nantes (France)} 
\small\it{$^{k}$GSI, Darmstadt (Germany)}
\small\it{$^{l}$NIKHEF, Utrecht (The Netherlands)}
\small\it{$^{m}$Univ. of Lund (Sweden)}
\small\it{$^{n}$Univ. of Tsukuba (Japan)}
\small\it{$^{o}$NPI, Rez (Czech Rep.)}
\small\it{$^{p}$KVI, Groningen (The Netherlands)}
\small\it{$^{q}$INS, Warsaw (Poland)}
\small\it{$^{r}$MIT, Cambridge (USA)}
\small\it{$^{s}$IOP, Bhubaneswar (India)}
\small\it{$^{t}$Univ. of Tennessee (USA)}

\newpage

\small
\begin{abstract} Abstract:~~~~
          Results from the multiplicity distributions of 
          inclusive photons and charged particles,
          scaling of particle multiplicities,
          event-by-event multiplicity fluctuations,
          and charged-neutral fluctuations 
          in 158$\cdot A$ GeV Pb+Pb collisions are
          presented and discussed. A scaling of charged particle
          multiplicity as $N_{part}^{1.07\pm 0.05}$ and
          photons as $N_{part}^{1.12\pm 0.03}$ have been observed,
          indicating violation of naive wounded nucleon model.
          The analysis of localized charged-neutral fluctuation
          indicates a model-independent demonstration of non-statistical
          fluctuations in both charged particles and photons in limited
          azimuthal regions. However, no correlated charged-neutral 
          fluctuations are observed.
\end{abstract}

\normalsize

\section{Introduction}
The year 2000 has been quite interesting and remarkable for heavy ion
physicists searching for a new form of matter, the quark-gluon 
plasma (QGP). In the beginning of the year, CERN has announced 
that there is definite evidence of possible QGP signals
from the experiments performed at the SPS \cite{qm99}.
This declaration was made after scrutinizing results from all the
dedicated experiments which have been taking data since 1994 using
the Pb beams at 158 GeV/nucleon (a total energy of 33 TeV).
In the middle of the year, this field entered a new era with the
commissioning of the Relativistic Heavy Ion Collider (RHIC) at
BNL which is a captive machine for QGP search. Soon after, new results
from the collisions at RHIC have started to appear \cite{qm01}. 
At this time, it is most appropriate to review one of the basic
observables of the nuclear collisions, that is the particle 
multiplicity distributions. One of the distinct advantages of the 
heavy ion collisions at such high energies is the production of large 
number of particles in every event. This allows for a detailed
study of event-by-event fluctuations in particle multiplicities and
ratios of identified particles. This is important for
understanding the evolution of nuclear system at high energy collisions.
Formation of disoriented chiral
condesates (DCC), which is a direct consequence of chiral symmetry
restoration, would also give rise to event-by-event
correlated fluctuations in charged particles to neutrals. In this
manuscript we discuss these topics in view of the data from the
WA98 experiment at the CERN-SPS.

For a thermalized 
system undergoing a phase transition, the variation of the temperature 
with entropy density is interesting as the temperature is expected 
to increase below the transition, remain constant during the 
transition, and then increase again \cite{shuryak,vanhove}.
This can be studied by measuring the mean transverse momentum,
$\langle p_T\rangle$, and particle multiplicities for
varying centrality, for a number of colliding systems at
different energies.
Since multiplicity of produced particles is an important
quantity to characterize the evolving system, its fluctuation from
event to event may provide a distinct signal of the 
phase transition from hadronic gas to QGP phase. 
Recently, several new methods using event-by-event fluctuations
of $\langle p_T\rangle$ and particle multiplicities
have been proposed to probe into the mechanism 
of phase transition \cite{step,heisel,asakawa,jeon}.
Another interesting phenomenon is the formation of domains
of DCC \cite{anselm,bj,blaizot,raj} which gives rise to isospin
fluctuation, in which the neutral pion fraction can deviate
significantly from 1/3, the value for uncorrelated emission of 
pions. Several methods have been proposed
to search for signals of DCC \cite{huang,qm97nayak,dccflow,dccstr,randrup}.
The most direct signal comes from the event-by-event fluctuations in the
number of charged particles to photons in localized
($\eta$-$\phi$) phase space.

The observed fluctuations will have contributions from
statistical fluctuations and those which have dynamical origin.
The contribution from dynamical origin comprise of (a) fluctuations
which do not change event-by-event, e.g., those from
Bose-Einstein (BE) correlations, resonance decays and (b) 
the fluctuations which have new physics origin and may
vary from event-to-event. The major interest for us would be to
probe into the fluctuations which have new physics origin, such as
those arising near the tricritical point of QCD phase diagram and
the formation of DCC. 
To extract the dynamical part from the observed fluctuations,
one has to understand the contributions from statistical and other
known sources. 
It is possible to probe at the non-statistical fluctuations 
from experimental data in a model independent way by comparing 
these with mixed events generated from the data. Once properly
understood, mixed events provide the best means to
infer about the presence of non-statistical fluctuations.

This manuscript is organized as follows. In the next section, we discuss
multiplicities of charged particles and photons and their pseudorapidity
distributions. In section~3 we discuss the  scaling of particle
multiplicities with centrality of the reaction expressed in terms of
the number of participants. In section~4 we present multiplicity
fluctuations and discuss the importance of making proper centrality
selection. In section~5 we discuss 
charged-neutral fluctuations in the full phase space
of the multiplicity detectors. In section~6 we present the
results of an analysis to search for localized fluctuations in smaller 
$\eta$-$\phi$ bins by comparing with several kinds of mixed events.
We give a summary in section~7.

\section{Multiplicity Distributions}

\begin{figure}
\setlength{\unitlength}{1mm}
\begin{picture}(130,50)
\put(10,5){
\epsfxsize=6.1cm
\epsfbox{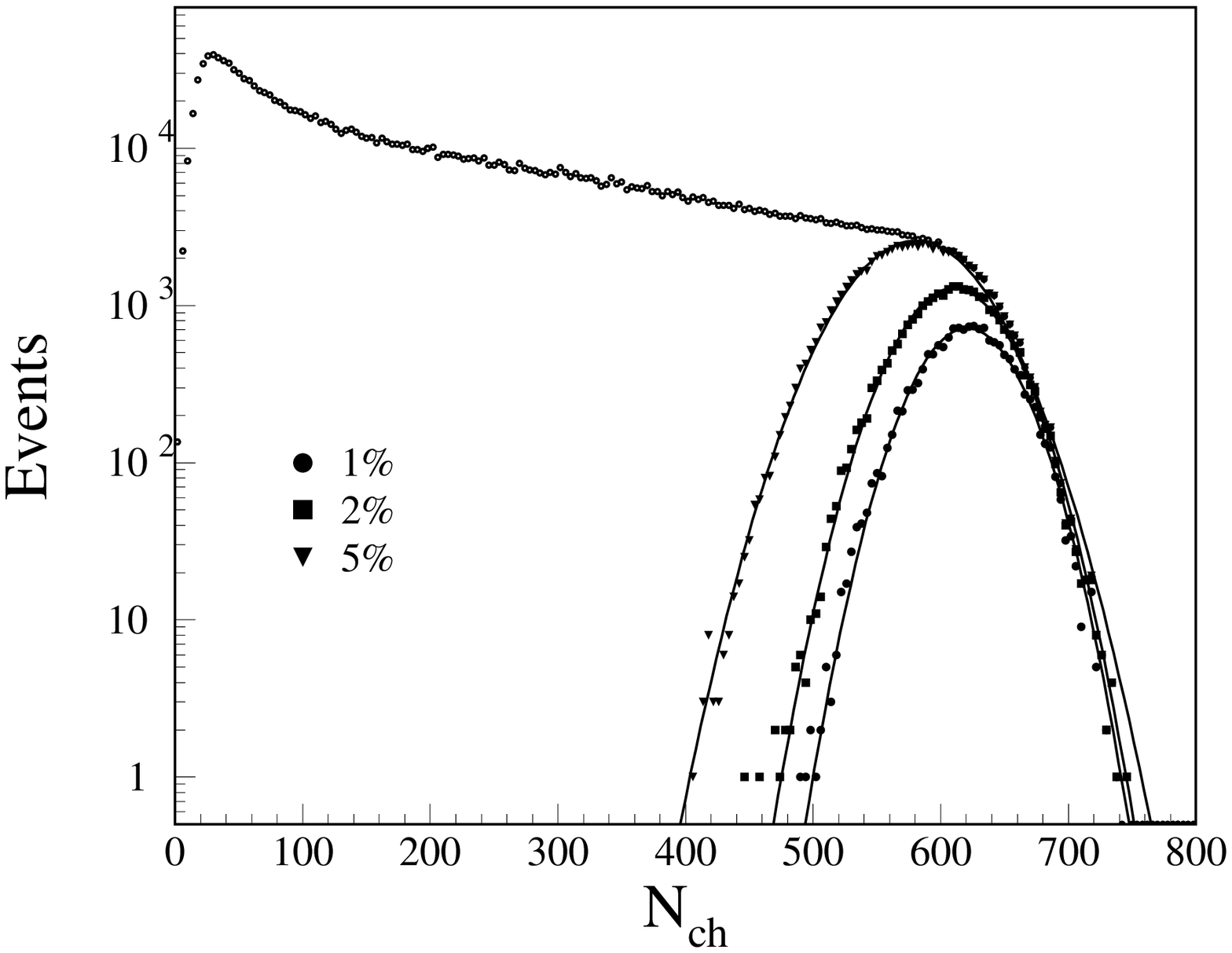}
}
\put(70,5){
\epsfxsize=6.1cm
\epsfbox{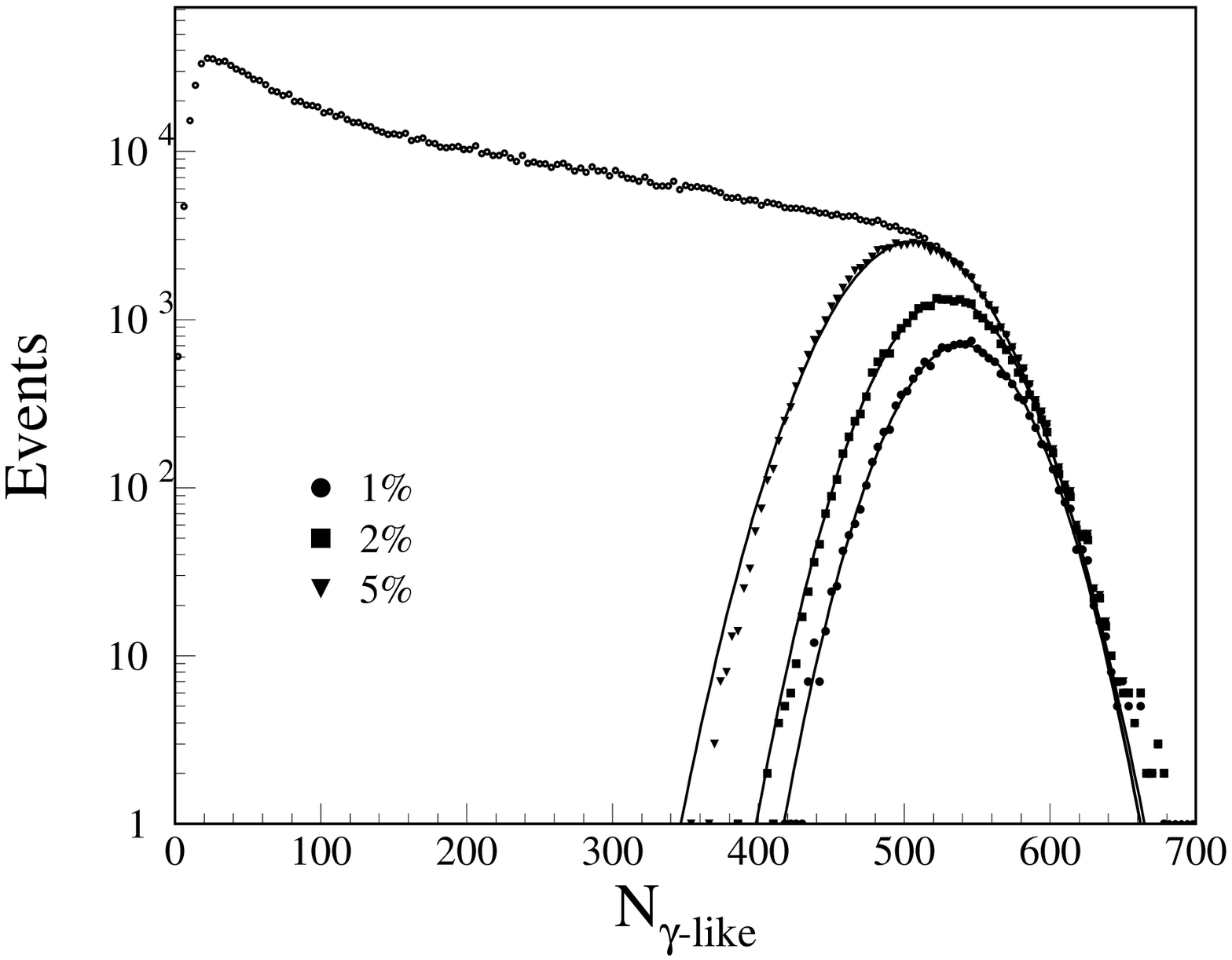}
}
\end{picture}
\vspace*{-1cm}
\caption{
Minimum bias multiplicity distributions of
charged particle and photon-like clusters for Pb+Pb reactions
at 158$\cdot A$ GeV. The other three curves
show multiplicity distributions for three different centralities,
0-1\%, 0-2\% and 0-5\%.}
\label{nch_ngam_cs}
\end{figure}

The WA98 experimental setup consists of large acceptance hadron and photon
spectrometers, detectors for charged particle and photon multiplicity 
measurements and calorimeters for measuring transverse and forward energies.
The charged particle hits were counted using a circular silicon pad
multiplicity detector (SPMD) located 32.8 cm from the target covering
$2.35<\eta<3.75$. The efficiency of detecting charged particles has
been determined in a test beam to be 99\%. The detector is transparent
to high energy photons. The photon multiplicity was measured using a
preshower Photon Multiplicity Detector (PMD) \cite{wa98nim} placed
21.5 meters downstream from the target and covering the pseudorapidity
range $2.9<\eta<4.2$. The cluster of hit pads, having total ADC content 
above
a hadron rejection threshold are identified as photon-like 
($N_{\gamma-{\rm like}}$).
The photon counting efficiencies for central to peripheral cases vary
from 68\% to 73\% \cite{WA98-9,wa98nim}. 
The purity of the photon sample in the two
cases varies from 65\% to 54\%.
The centrality of the interaction is determined by the total
transverse energy ($E_{\rm T}$) measured in the mid rapidity calorimeter.
The centralities are expressed as fractions of the measured total 
transverse energy. The most central corresponds to the top 5\% of the
minimum bias cross section. Extreme peripheral events in the 80-100\% range
were not analyzed.

Fig.~\ref{nch_ngam_cs} shows the minimum bias charged particle
and photon-like distributions within the full acceptance of
the detectors. The distributions corresponding to the centrality
cuts of 1\%, 2\% and 5\% of minimum bias cross section are superimposed
on the figure.

\begin{figure}
\setlength{\unitlength}{1mm}
\begin{picture}(130,45)
\put(10,5){
\epsfxsize=5.5cm
\epsfbox{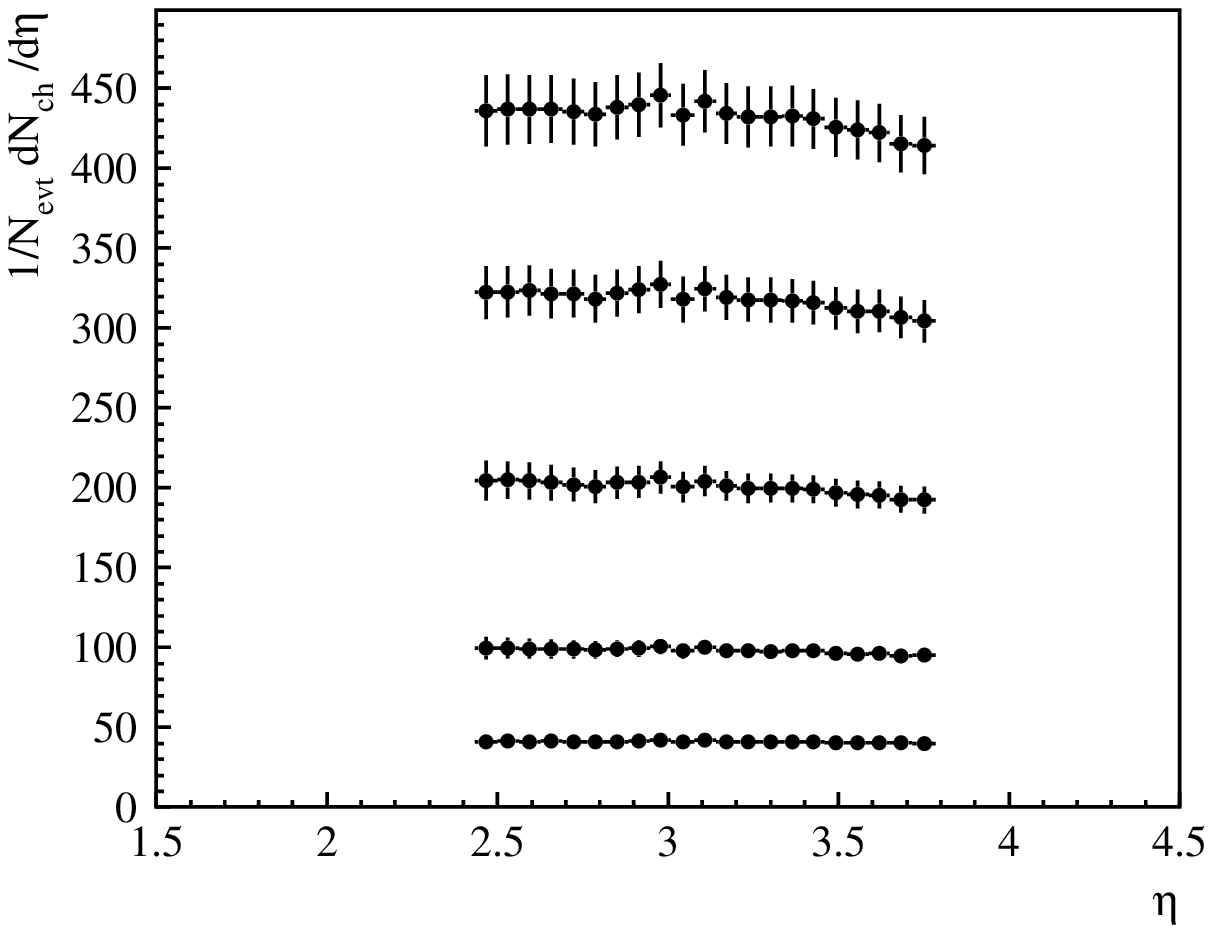}
}
\put(63,2){
\epsfxsize=6.1cm
\epsfbox{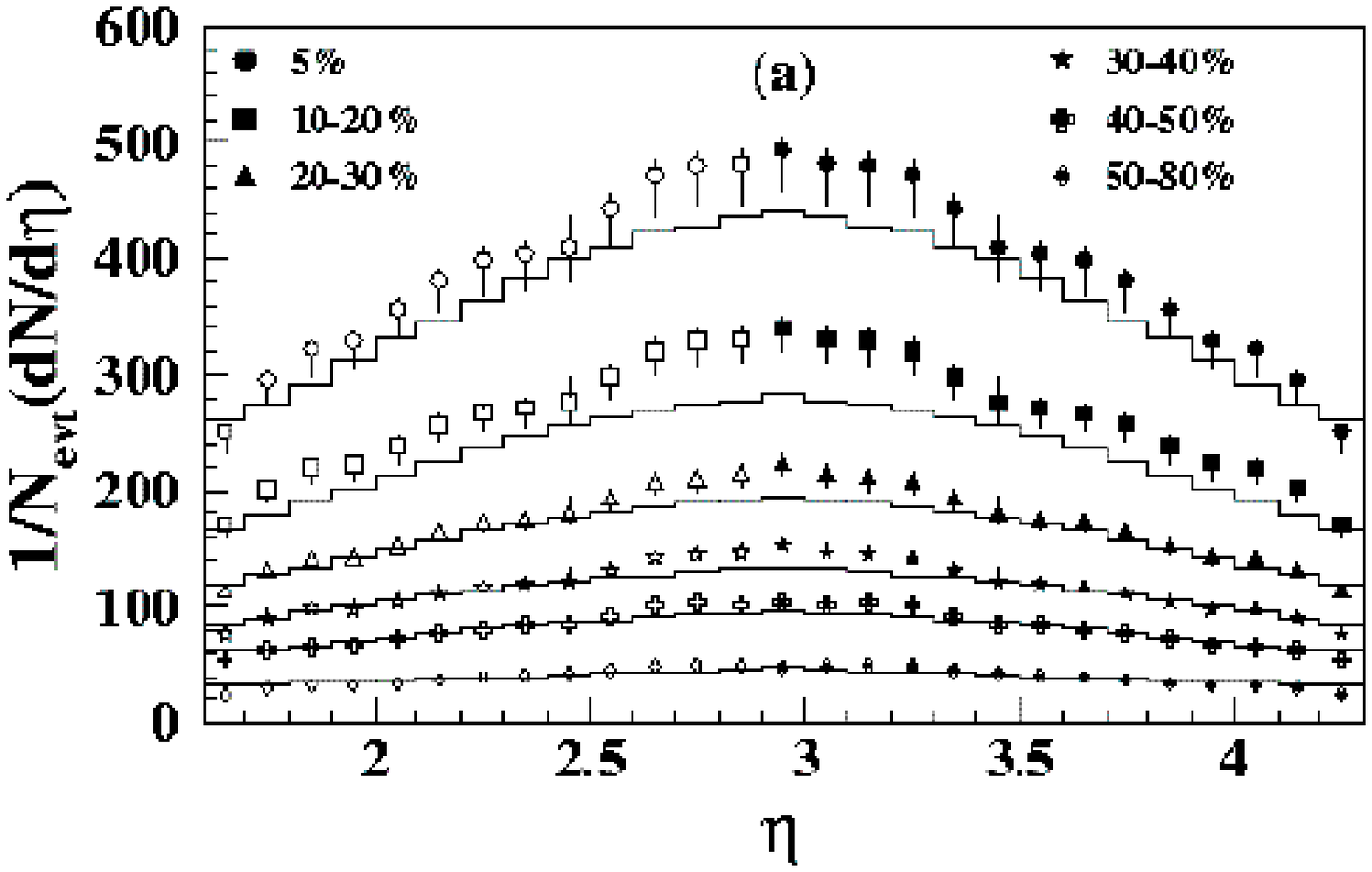}
}
\end{picture}
\vspace*{-1cm}
\caption{
Pseudorapidity distributions of charged particles and photons 
in Pb induced reactions at 158$\cdot A$ GeV on a Pb target.
The charged particle distributions shown in the left panel are
for 0-10\%, 10-20\%, 20-40\%, 40-60\% and 60-80\% centrality bins.
The centrality bins for photons are as labelled in the figures.
}
\label{nch_ngam_eta}
\end{figure}

    The pseudorapidity distribution of charged particles and
    photons at different centralities are shown in the
    left and right panels of Fig.~\ref{nch_ngam_eta}, respectively.
    The data have been corrected for geometry and efficiency factors. 
    For photons, the filled symbols represent
    the measured data, and the open symbols are reflections of the
    filled symbols at $\eta_{c.m.}(=2.92)$.
    The solid histograms show the corresponding distributions obtained from
    VENUS event generator \cite{venus}.

\section{Scaling of particle production}

The gross features of particle production in nucleon-nucleon collisions and
reactions of light nuclei are well described in the framework of wounded
nuclear model \cite{bialas}. In this model the transverse energy 
and particle
production in p+A and A+A reactions are calculated by assuming a constant 
contribution from each participating nucleon. While a scaling with 
the number
of collisions arises naturally in a picture of a superposition of 
nucleon-nucleon collisions, with a possible modification by initial state 
effects, the scaling using number of participants is more 
naturally related to a
system with strong final state rescattering, where the incoming 
particles loose
their memory and every participant contributes a similar amount of energy
to particle production. The scaling behavior of particle production may 
therefore carry important information 
on the reaction dynamics \cite{WA98-10}.

\begin{figure}
\setlength{\unitlength}{1mm}
\begin{picture}(130,55)
\put(7,5){
\epsfxsize=6.3cm
\epsfysize=6.3cm
\epsfbox{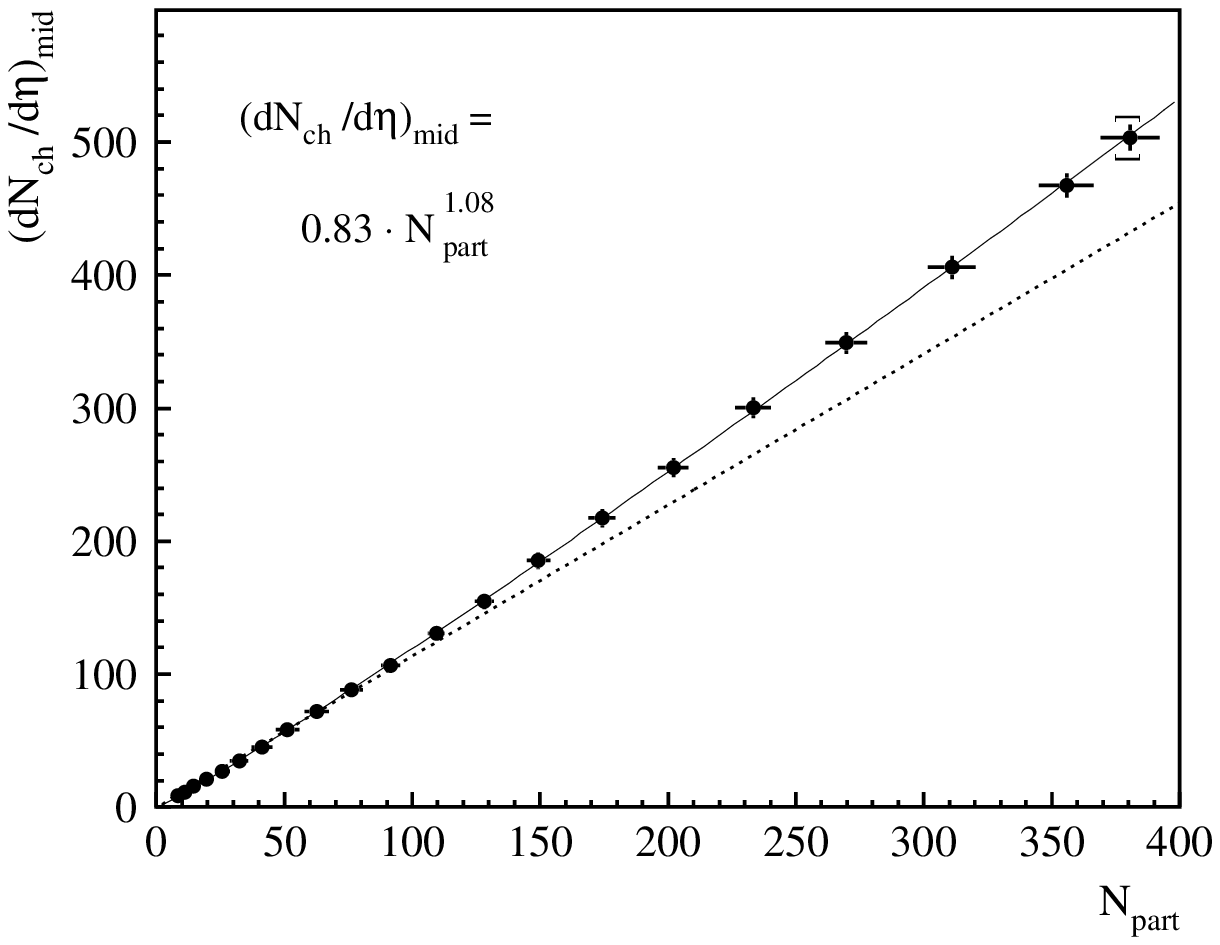}
}
\put(68,2){
\epsfxsize=6.1cm
\epsfysize=6.5cm
\epsfbox{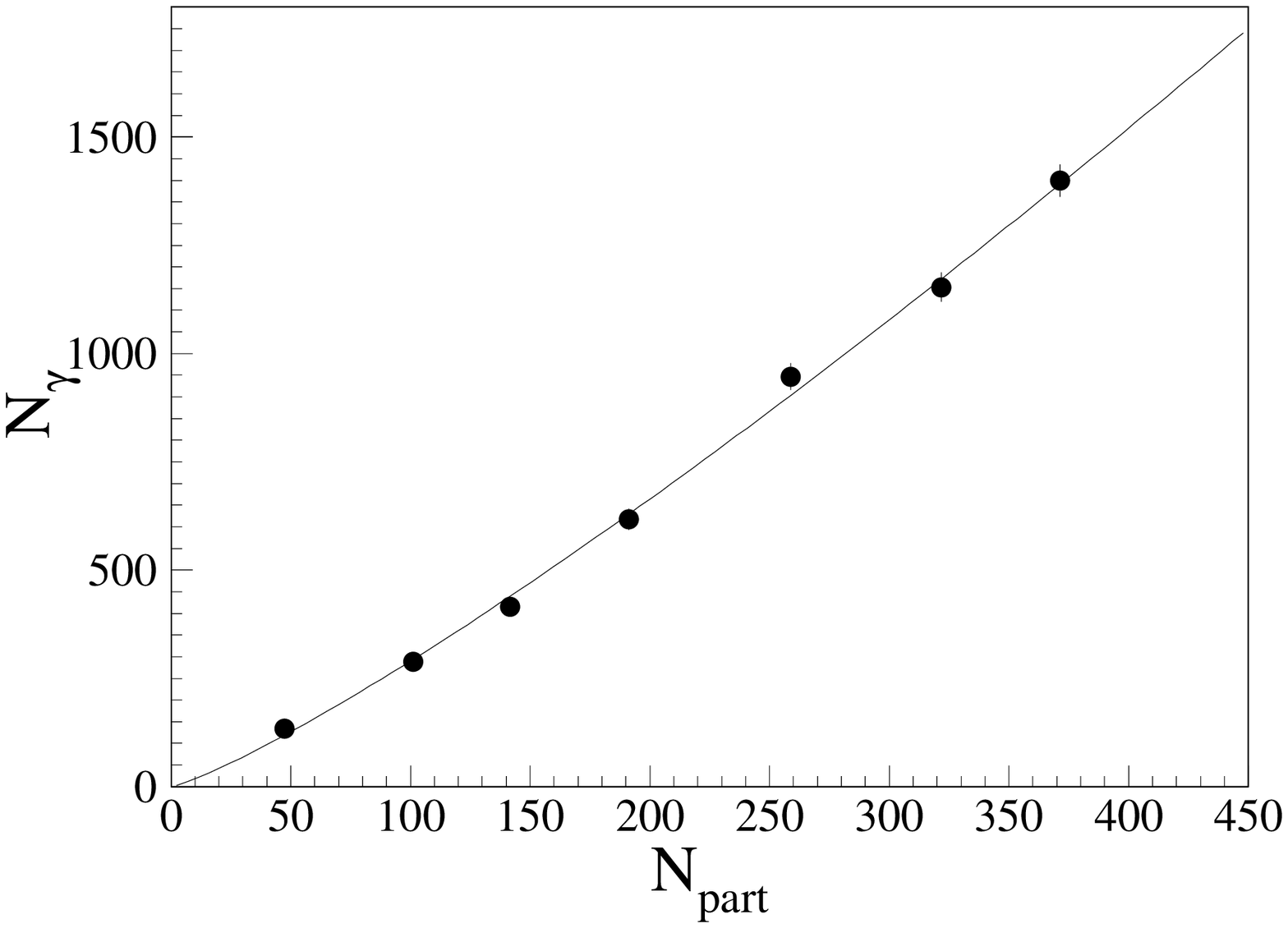}
}
\end{picture}
\vspace*{-1cm}
\caption{
    Scaling behavior of charged particles and photons. Pseudorapidity
    density of charged particles at midrapidity and integrated number
    of photons are plotted as functions of number of participating
    nucleons.
    The solid lines show power-law fit to the data,
    which yields the value of the exponent, $\alpha=1.07\pm 0.05$ for charged
    particles and $1.12\pm 0.03$ for photons, respectively.
}
\label{nch_ngam_scale}
\end{figure}

The number of participants have been calculated using the 
VENUS model. Pseudorapidity density of $N_{\rm ch}$ 
at midrapidity as a function of the number
of participants for Pb+Pb collisions are shown in the left panel of 
Fig.~\ref{nch_ngam_scale}. The data points show stronger than linear
increase (shown as dotted line). A scaling relation can be obtained
by fitting the data points using $C\times N_{part}^\alpha$. The value
of $\alpha$ is extracted to be $1.07\pm 0.05$ \cite{WA98-10}. 
Similarly on the right side
we have shown the integrated number of photons as a function of number
of participating nucleons. A similar power law fit yields the value
of $\alpha$ to be $1.12\pm 0.03$\cite{WA98-9}. The results for
charged particles and photons are consistent with each other within
the quoted errors.
These results show that there is a clear
participant scaling violation compared to a purely linear dependence.
The scaling violation might have consequences on many other signals,
for example, on the $J/\psi$ production.

\section{Multiplicity Fluctuations}

A lot of theoretical interest has been generated on the subject of
event-by-event fluctuations, primarily motivated by the near perfect
Gaussian distributions of several observables for a given centrality
bin. If the distribution of a quantity $X$ is Gaussian, then
one defines the amount of fluctuation by the following:
\begin{equation}
\omega_X = \frac{\sigma_X^2}{<X>}
\end{equation}
where $\sigma_X$ is the variance of the distribution. That is, the fluctuation
of the distribution is the variance squared normalized to the mean of the
distribution under consideration. 

\begin{figure}
\setlength{\unitlength}{1mm}
\begin{picture}(130,55)
\put(10,0){
\epsfxsize=5.0cm
\epsfbox{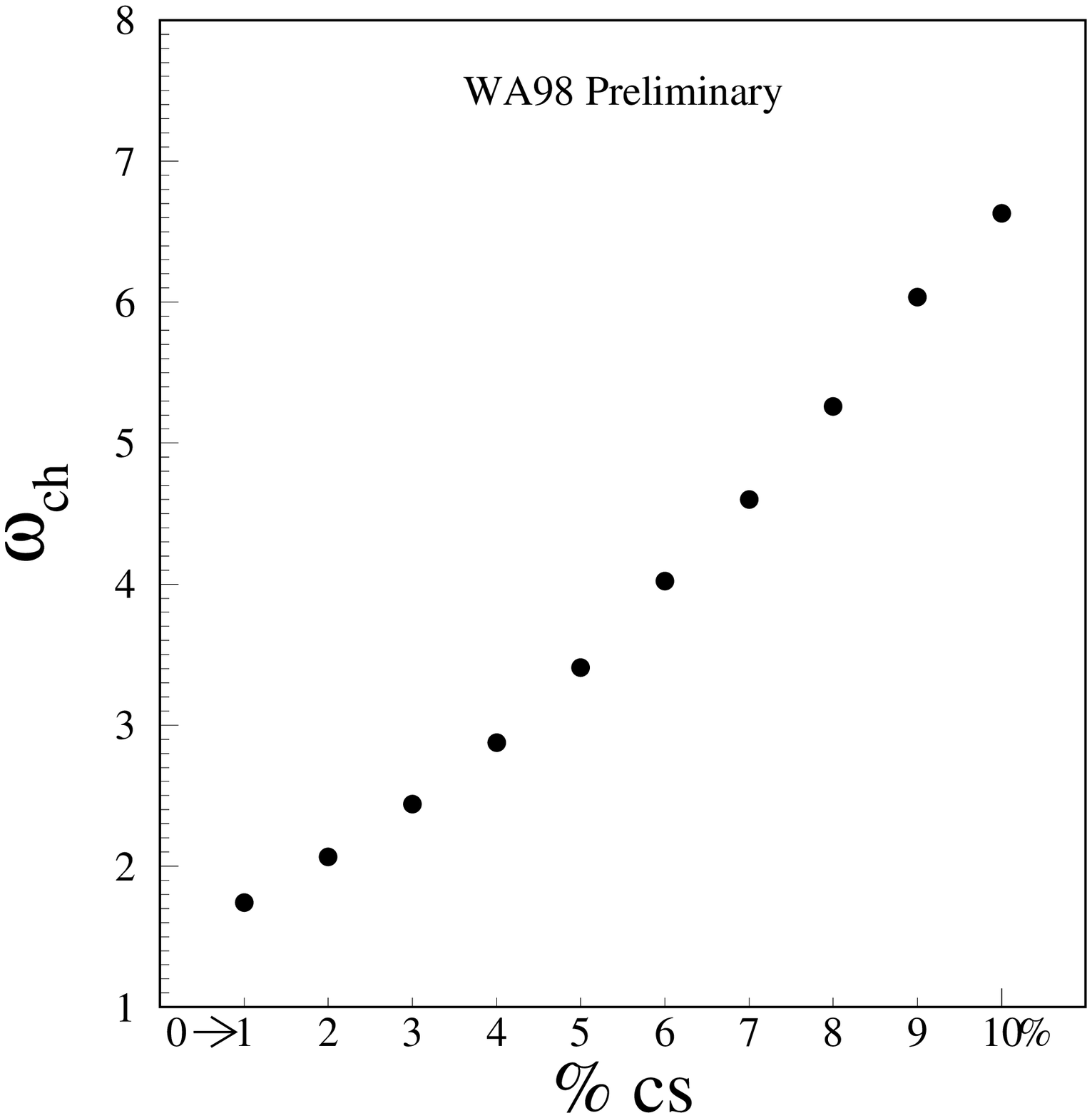}
}
\put(70,0){
\epsfxsize=5.0cm
\epsfbox{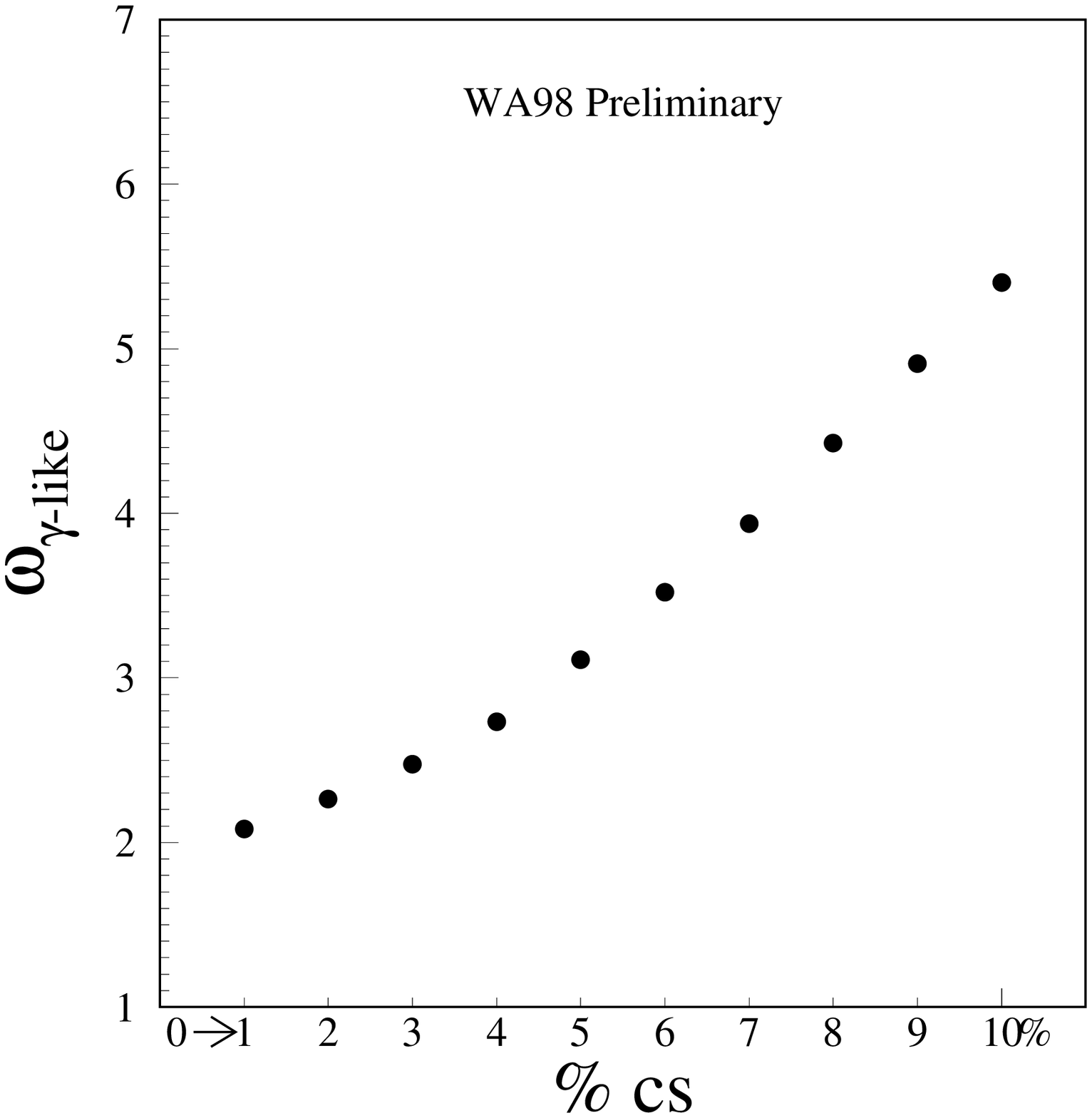}
}
\end{picture}
\bigskip
\vspace*{-1cm}
\caption{Multiplicity fluctuations of charged particles and photons for
various centralities within the full coverage of the PMD and SPMD.
The width of the centrality bins as percentage of minimum bias cross section
increases along the x-axis.
}
\label{nch_ngam_w_cs}
\end{figure}

In Fig.~\ref{nch_ngam_cs}, the distributions of $N_{\rm ch}$ and 
$N_{\gamma-{\rm like}}$ are shown for centrality bins of 
0-1\%, 0-2\% and 0-5\%. These curves are very good Gaussians with
fits giving $\chi^2$/ndf to be close to unity. It has been observed
that making the centrality bin broader beyond 0-5\%,
(from 0-6\% and beyond) the distributions deviate from Gaussians.
Using the mean and variance of the distributions at different
centrality bins we have calculated the amount of fluctuations using
eqn. (1). These results are plotted in Fig.~\ref{nch_ngam_w_cs} for
charged particles and photons. It is seen that the amount of
fluctuation increases by increasing the width of the centrality
bin. This is obvious from the fact that by increasing the width of
the centrality selection bin, the inherent statistical fluctuations
also increases. 

\begin{figure}
\setlength{\unitlength}{1mm}
\begin{picture}(130,55)
\put(10,0){
\epsfxsize=5.0cm
\epsfbox{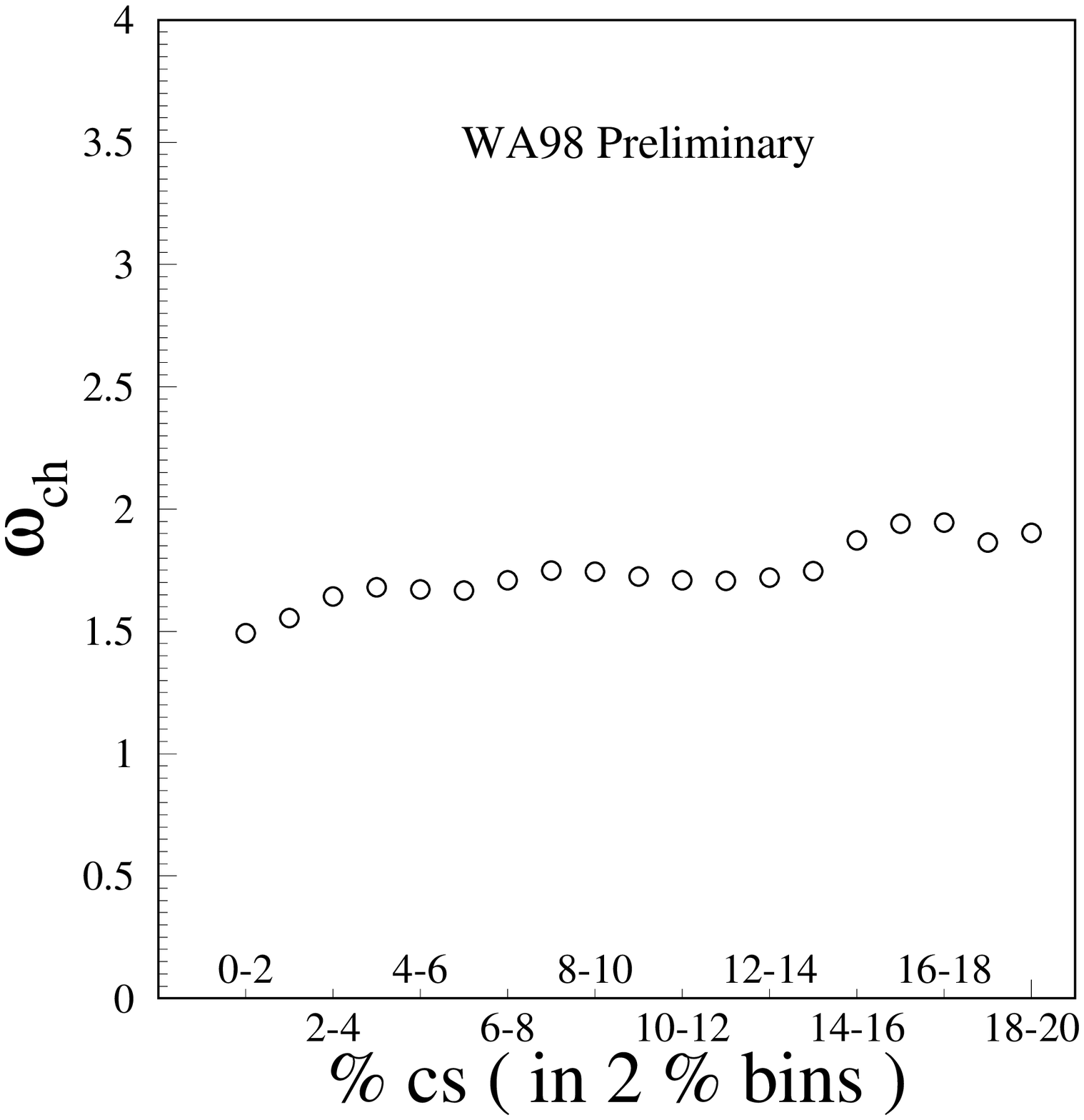}
}
\put(70,0){
\epsfxsize=5.0cm
\epsfbox{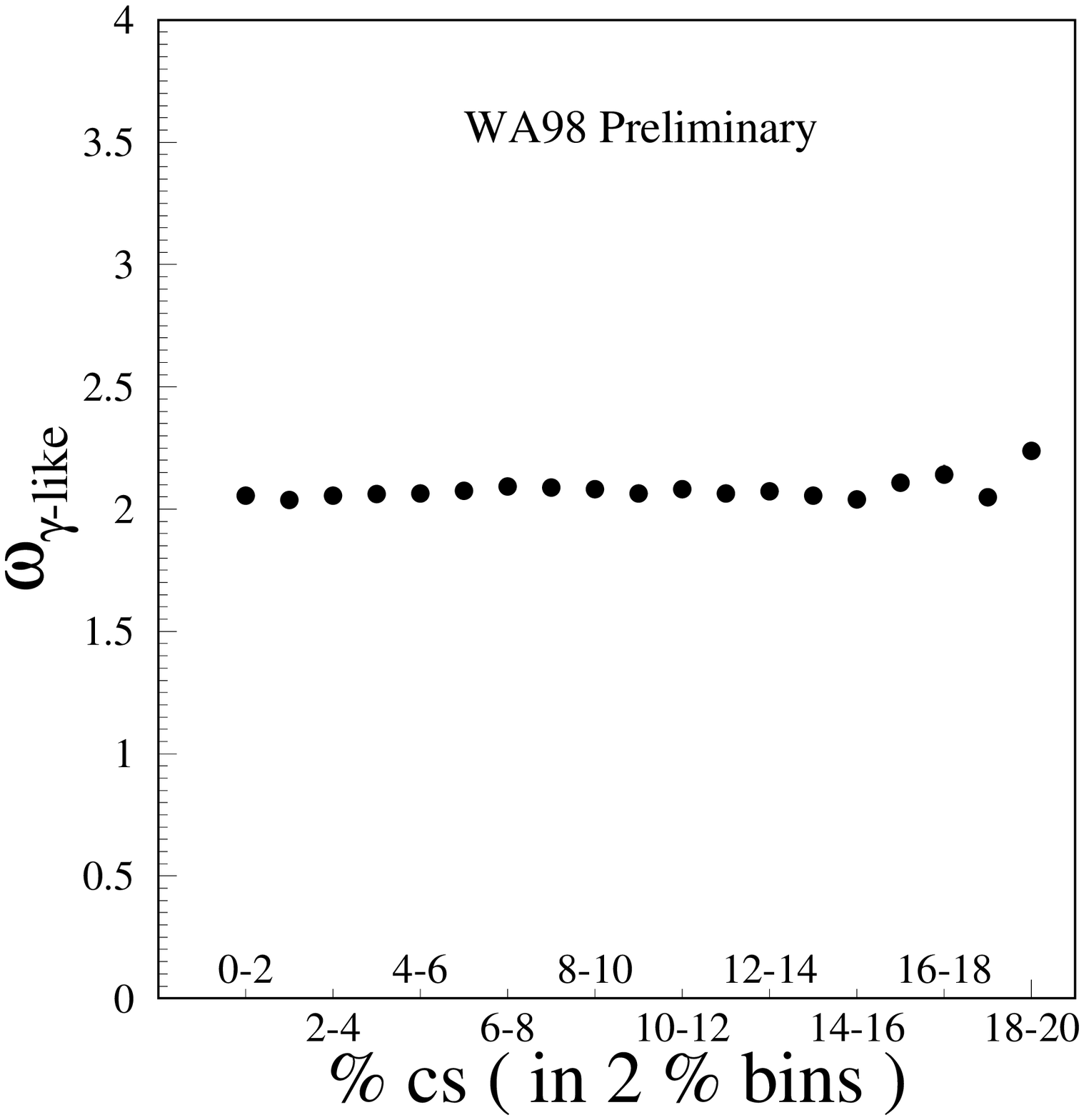}
}
\end{picture}
\bigskip
\vspace*{-1cm}
\caption{Multiplicity fluctuations of charged particles and photons for
various centralities in a common coverage of the PMD and SPMD.
The width of the centrality bins as percentage of minimum bias cross section
remain same (2\%) along the x-axis.
}
\label{nch_ngam_w}
\end{figure}

Thus it is very important to control centrality properly
in all fluctuation studies such that the multiplicity distributions
are good Gaussians. Keeping this aspect in mind we have used a different 
set of centrality selection by taking 2\% cross section bins,
such as, 0-2\%, 2-4\%, 4-6\%, 6-8\%, etc. 
The resulting multiplicity distributions are good Gaussians in
nature with $\chi^2$/ndf between 1 and 1.5. The amount of
fluctuations are calculated for these type of centrality bins and plotted
in Fig.~\ref{nch_ngam_w}. The fluctuations for charged particles
increase weakly by going from highest centrality towards peripheral whereas
for photons the fluctuations are almost uniform. The fluctuations in 
photons are higher compared to those of the charged particles. This
difference may arise because majority of photons are decay products of
$\pi^0$.
These results have to be put in perspective in terms of the contributions
from known sources, such as, effect of finite multiplicity, fluctuations
because of impact parameter of the reaction, effect of rescattering,
BE correlations, and resonance decays.

\section{Charged-neutral fluctuations}

We now turn to event-by-event fluctuations in charged particle and
photon multiplicities. We make a global event characterization in 
terms of the photon and charged particle distributions over the full
available phase space of the detectors. The main motivation is to 
search for single large size DCC domains by looking for events which 
fall far beyond the correlation line of $N_{\gamma-{\rm like}}-N_{\rm ch}$
distributions.

\begin{figure}
\setlength{\unitlength}{1mm}
%\vspace{-0.4cm}
\begin{picture}(140,60)
\put(0,0){
\epsfxsize=5.5cm
\epsfysize=5.0cm
\epsfbox{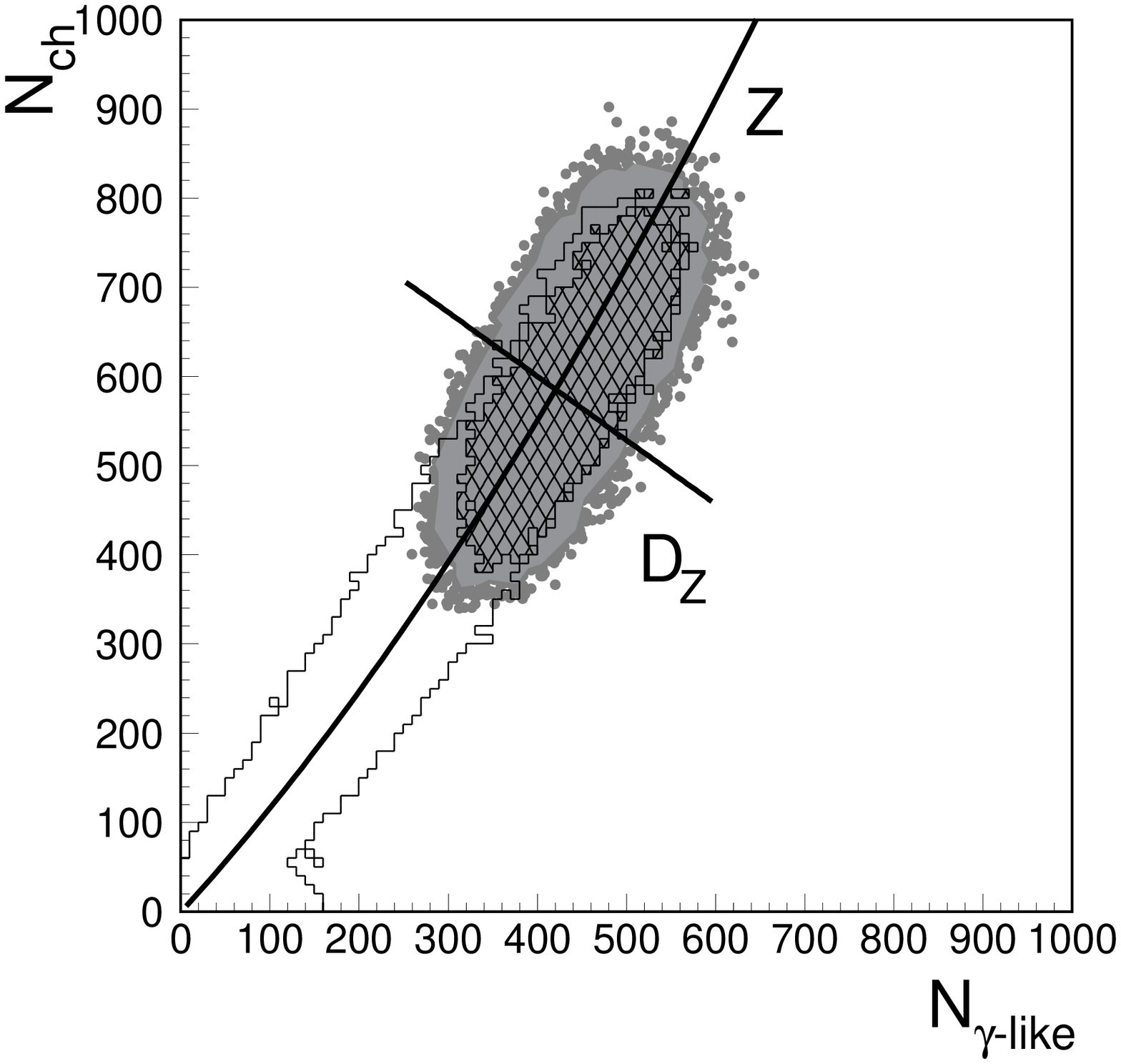}
}
\put(70,0){
\epsfxsize=5.5cm
\epsfysize=5.0cm
\epsfbox{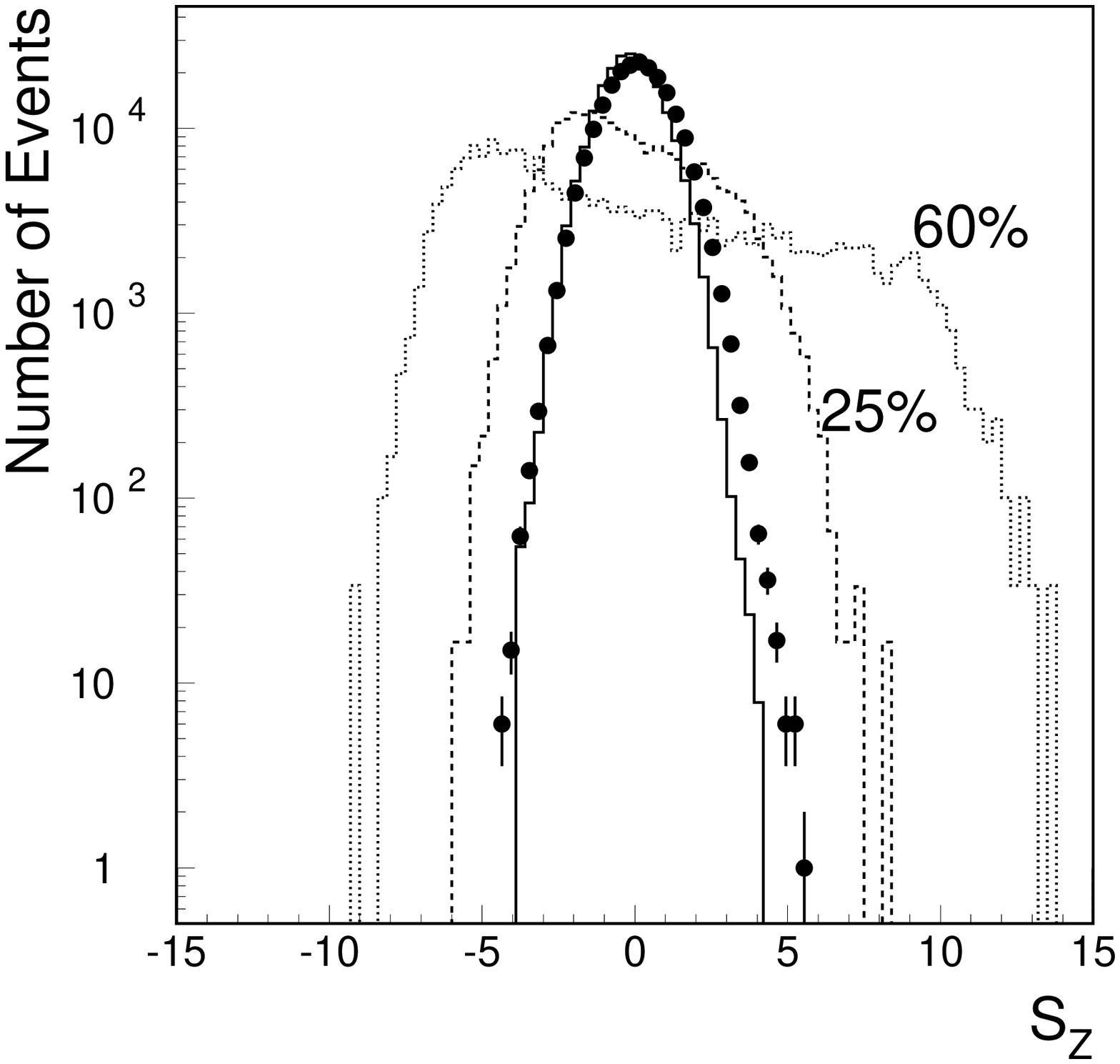}
}
\end{picture}
\vspace{-0.3cm}
\vspace*{-0.5cm}
\caption {\label{dcc_global}
The left panel shown the correlation of $N_{\gamma-{\rm{like}}}$ 
and $N_{ch}$ for central (top 10\%) events. The hatched region is for 
VENUS+GEANT simulation results. The right panel shows the 
$S_Z$ distribution for the experimental data is shown, 
overlaid with VENUS+GEANT results (solid histogram). The
other curves are generated by incorporating 25\% and 60\% DCC in every event.
}
\end{figure}

The correlation plot of $N_{\gamma-{\rm{like}}}$ 
and $N_{ch}$ for top 10\% central events is shown in the
left panel of Fig.~\ref{dcc_global}. 
A strong correlation is seen between charged and neutral multiplicities.
The fitted line of the distribution is represented by the $Z$ axis.
The closest distance ($D_{Z}$) of the data points to the
correlation axis has been calculated numerically with the
convention that $D_{Z}$ is positive for points below the Z-axis.
We have chosen to work with the scaled variable
$S_Z{\equiv}D_Z/\sigma_{D_Z}$ in order to compare relative fluctuations
at different multiplicities.
$S_Z$ distribution for the data is shown as filled circles 
in the right panel of Fig.~\ref{dcc_global}. the solid
histogram shows the results from the VENUS events passed
through the GEANT simulation package of WA98 experiment.
These events are termed as (V+G) in rest of the manuscript.
The DCC events are expected to show up as non-statistical
tails in the distribution of $S_Z$. This is illustrated by
the other curves where one introduces a known about of DCC
fluctuation. Introduction of DCC makes the curves broader.
Since we do not see any such event in our data sample,
we are faced with
the possibilities that single-domain DCCs are very rare, very small, or both.
Based on a simple DCC model calculation we have put upper limits
on the frequency of DCC production as a function of
its size. More details may be obtained from \cite{WA98-3}.
We discuss search for small size DCC in the next section.

\section{Localized charged-neutral fluctuations}

After studying the charged-neutral fluctuations
on the full available phase space, the next interest would
be to search for fluctuations in localized ($\eta$-$\phi$) phase space
regions or domains \cite{WA98-12}. One of the major interest 
would be the search
for the formation of small size DCC as it is supposed to produce clusters of
coherent pions in localized domains. The probability distribution
of neutral pion fraction in a DCC domain follows the relation :
\begin{equation}
P(f) = 1/2\sqrt{f} ~~~~~~~~~~~~~~~~{\rm where}~~~~~~~~~~
 f = N_{\pi^0}/N_{\pi}.
\end{equation}
Thus DCC formation in a given domain would be associated with
large correlated event-by-event fluctuations in the multiplicities of
charged particles and photons as majority of charged particles consist
of charged pions and majority of the photons originate from $\pi^0$ decays.

The present analysis uses data from top 5\% central events,
which corresponds to a total of 85K events. The pseudorapidity region
common to both PMD and SPMD is selected. The acceptance in terms of 
transverse momentum ($p_{\rm T}$) extends down to 30 MeV/c, although no explicit
selection in $p_{\rm T}$ is applied. 
The experimental results are compared to simulated V+G
events and various types of mixed events.

\subsection{A simple DCC model}

The effect of non-statistical DCC-like charged-neutral fluctuations
has been studied within the framework of a simple model in which
the output of the VENUS event generator has been modified.
In this, the charges of the pions within a localized $\eta$-$\phi$ 
region from VENUS are interchanged
pairwise ($\pi^{+}\pi^{-} \leftrightarrow \pi^{0}\pi^{0}$),
according to the DCC probability
distribution as given in equation (2). 
For the present study, DCC events have been generated in the
range of $3.0\le \eta \le 4.0$ with varying intervals in $\Delta\phi$.
An ensemble of events (henceforth referred to as nDCC events)
were then generated by mixing different fractions
of DCC-like events with those of normal events, 
as appropriate for different probabilities of occurrence of DCC.
After allowing the $\pi^0$s to decay, all
the particles were then tracked through the full GEANT simulation
program of WA98 experiment. These events are then analyzed using
the same analysis methods as of the data.

Two different analysis methods have been used for the search of
non-statistical fluctuations. The first one is the method of
discrete wavelet transformation (DWT) and the second one is the
$N_{\gamma-{\rm like}-N_{\rm ch}}$ correlation.

\subsection{Discrete wavelet transformation}

The DWT method \cite{numeric} has been 
utilized very successfully in many fields including image processing,
data compression, turbulence, human vision, radar,
and earthquake prediction \cite{amara}.
The beauty of the DWT technique lies in its ability to
analyze a spectrum at different resolutions with the ability to
pick up any fluctuation present at the right scale.
Simulation studies by Huang et. al. \cite{huang} and
Nandi et. al. \cite{dccstr} have shown that
the DWT analysis could be a powerful technique for the search of
localized DCC.

While there are several families of wavelet bases distinguished by
the number of coefficients and the level of iteration, we have used
the $D$-4 basis \cite{numeric}. The analysis has been performed with the sample
function,
\begin{equation}
        f^\prime = \frac {N_{\gamma-{\mathrm like}}}
                              {N_{\gamma-{\mathrm like}}+N_{\mathrm ch}},
\end{equation}
calculated at the highest resolution scale, $j_{max}=5$. The spectrum
of $f^\prime$ is the input to the DWT analysis.
The sample function is then analyzed at different scales $j$ by
being re-binned into 2$^j$ bins.
The output of the DWT consists of a set of wavelet or father function
coefficients (FFCs) at each scale, from $j=1$,...,($j_{max}-1)$.
The coefficients obtained at a given scale, $j$, are derived from the
distribution of the sample function at one higher scale, $j+1$. The
FFCs quantify the deviation of the bin-to-bin fluctuations in the
sample function at that higher scale relative to the average behavior.
The distribution of FFCs for normal events
is Gaussian in nature. However, the
presence of localized non-statistical fluctuations makes the distribution
broader, with a larger rms deviation of the 
FFC distribution \cite{huang,dccflow,dccstr}. 
Comparing the rms deviations of the FFC distribution
of data and mixed events one can get an idea about the
localized fluctuations.

\subsection{$N_\gamma-{\rm like}$ and $N_{\mathrm ch}$ correlations}

\begin{figure}
\epsfxsize=7cm
\centerline{\epsfbox{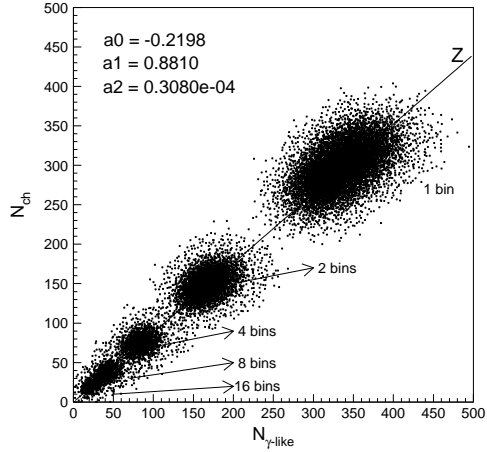}}
\vspace*{-1cm}
\caption {\label{data_dz_2d}
The scatter plot showing the correlation of
$N_{\gamma-{\mathrm like}}$ and $N_{\mathrm {ch}}$ for 1,2,4,8 and 16 bins in
azimuthal angle, $\phi$, for a common coverage of the two detectors,
PMD and SPMD. 
The data are for central events, corresponding to top
5\% of the minimum bias cross section. The scatter
plots at different bins are fitted to a curve (shown as Z axis) with 
a second order polynomial. The fit parameters are shown.
}
\end{figure}

   The correlation between $N_{\gamma-{\rm like}}$
   and $N_{\rm ch}$ has been studied 
   in smaller $\phi$-segments by dividing the
   $\phi$-space into 2, 4, 8 and 16 bins.
   The correlation plots of $N_{\gamma-{\rm like}}$ and $N_{\rm ch}$ have been
   constructed for each $\phi$ bin. This is shown in Fig.~\ref{data_dz_2d}.
   A common correlation axis ($Z$) has been obtained by fitting
   the above distributions with a second order polynomial.
   The closest distance ($D_{Z}$) of the data points to the
   correlation axis has been calculated numerically with the
   convention that $D_{Z}$ is positive for points below the 
   Z-axis.
   The distribution of $D_Z$ represents the
   relative fluctuations of $N_{\gamma-{\rm like}}$ and $N_{\rm ch}$ 
   from the correlation axis at any given $\phi$ bin.
   In order to compare these fluctuations at different scales in the
   same level, we work with a scaled variable,
   $S_{Z} = D_Z/{\mathrm rms}(D_Z)$, where ${\mathrm rms}(D_Z)$ corresponds
   to VENUS events. The presence of events with localized
   fluctuations in $N_{\gamma-{\mathrm like}}$ and $N_{\mathrm ch}$, 
   at a given $\Delta\phi$ bin, 
   is expected to result in a broader distribution of $S_Z$ compared
   to those for normal events at that particular bin. 

\subsection{Mixed events}

    In order to search for the presence of fluctuations
    in the experimental data, it is necessary to understand all detector
    related effects by generating different sets of mixed events,
    keeping specific physics goals in mind.
    Properly constructed mixed events should preserve all detector effects
    while removing correlations. Four sets of mixed events
    are generated to provide equivalent event samples as real events.
    In each type of mixed event the global (bin 1) 
    $N_{\gamma-{\rm like}}-N_{\rm ch}$ correlation is maintained as in the
    real event.

    The first type of mixed events (M1) are generated by mixing hits on
    both the PMD and SPMD hits separately, with no two hits
    taken from the same detector. Hits within a detector in the mixed events
    are not allowed to lie within the two track resolution of that detector.
    The second kind of mixed events (M2) are generated by mixing unaltered
    PMD hits of one event with the unaltered SPMD hits of a different event.
    Two more mixed event types are possible which are intermediate between 
    the M1 and M2 mixed events: (i) the hits within the PMD are 
    unaltered while the SPMD hits are mixed, this is called M3-$\gamma$,
    and (ii) the SPMD hits are unaltered while the PMD hits are mixed,
    this is called M3-ch. In table~1 we summarise the construction
    and usefulness of each of the mixed events.

\begin{table}
\begin{center}
\caption{Types of mixed events and how they are used for
         different physics.}
\medskip
\begin{tabular}{|cccc|} \hline
             & PMD      & SPMD       & Type of fluctuation to probe: \\ \hline
M1           & Mix hits & Mix hits   & Correlated + Individual \\ 
M2           & Unaltered hits &  Unaltered hits   & Correlated \\ 
M3-$\gamma$  &  Unaltered hits &  Mix hits  & $N_{\gamma}$ only \\ 
M3-ch        &  Mix hits &    Unaltered hits  & $N_{\rm ch}$ only \\ \hline
\end{tabular}
\end{center}
\end{table}

\begin{figure}
\epsfxsize=8cm
\centerline{\epsfbox{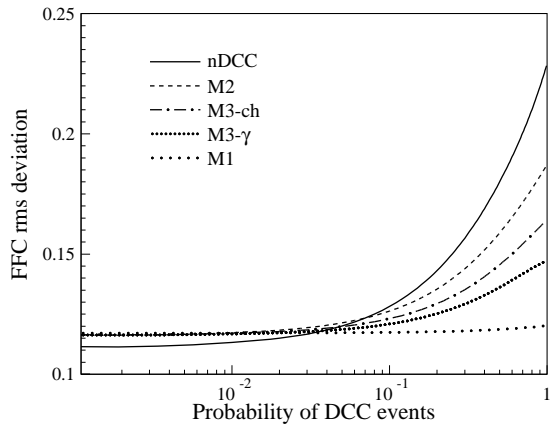}}
\vspace*{-1cm}
\caption {\label{m1_m2_check}
The rms deviations of the FFC distributions at $j=1$ for
simulated nDCC events with extent $\Delta\phi_{DCC}=90^\circ$ 
and for various mixed events constructed from those events, as
a function of the fraction of DCC-like events present 
in the nDCC sample.
}
\end{figure}

The behavior of the mixed events may be understood from the
sample of nDCC events. We have constructed all four different
kinds of mixed events from the nDCC events and analyzed using
both the methods. Here we give the results of the DWT analysis.
The rms deviations of the FFCs for nDCC events and the different kinds of 
mixed events produced from the nDCC events are shown in
Fig.~\ref{m1_m2_check}. In the absence of DCC-like fluctuation
the rms values of the various types of mixed events are very close to 
each other. But the V+G rms values come out to be lower than
those of the mixed events. This is 
due to the presence of additional correlations between
$N_{\mathrm ch}$ and the charged particle contamination in the 
$N_{\gamma-{\mathrm like}}$ sample.

The rms deviations for the M1-type of mixed events
are found to be almost independent 
of probability of DCC-like events. Thus it provides a baseline
for studying non-statistical fluctuations. The 
rms deviations of the M2 type of mixed
events increase similarly, but more weakly, than those of the
nDCC events. The rms deviations for
the M3 sets of events are found to lie between M2 and M1. Thus, the
sequence of the mixed events relative to the simulated nDCC events (or data)
gives a model independent indication of the presence and source
of non-statistical fluctuations. The simple DCC model used here 
results in an anti-correlation between $N_{\gamma-{\mathrm like}}$ and 
$N_{\mathrm ch}$ due to the ``isospin-flip'' procedure used to
implement the DCC effect. It also results in non-statistical fluctuations
in both $N_{\gamma-{\mathrm like}}$ and $N_{\mathrm ch}$. Thus the 
M2 mixed events remove only the $N_{\gamma-{\mathrm like}}$--$N_{\mathrm ch}$
anti-correlation while the
M1 mixed events are seen to remove all non-statistical fluctuations and
correlations. The M3 mixed events give intermediate results because
they contain only the $N_{\gamma-{\mathrm like}}$ (M3-$\gamma$) or
$N_{\mathrm ch}$ (M3-${\mathrm ch}$) non-statistical fluctuations.

\subsection{Results}

\begin{figure}
\epsfxsize=8cm
\centerline{\epsfbox{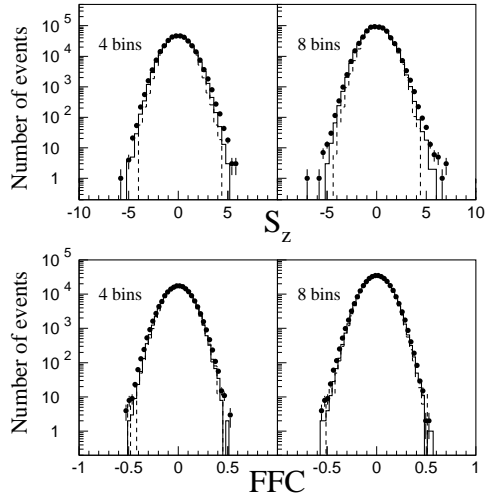}}
\vspace*{-1cm}
\caption {\label{sz-ffc}
The $S_Z$ and FFC distributions for 4 and 8 divisions in $\phi$.
The experimental data, M1 and V+G events are
shown by solid circles, solid histograms and dashed histograms,
respectively.
The statistics for data and mixed events are the same, whereas the
distribution for the V+G events is normalized to the number of data
events.
}
\end{figure}

The $S_{Z}$ distributions  
calculated at 4 and 8 bins in $\phi$ angle are 
shown in the top panel of Fig.~\ref{sz-ffc} for data, M1
and V+G events. The experimental data is broader than
the simulation and M1 events, indicating the presence of 
additional fluctuations.
The FFC distributions extracted from the measured $f^\prime(\phi)$
ratio are shown in the bottom panel of Fig.~\ref{sz-ffc} 
for the experimental data, for M1-type mixed events (from data), 
and for V+G events. The results are shown for scales $j=$1 and 2,
which carry information 
about fluctuations at $90^\circ$ and $45^\circ$ in $\phi$, 
The FFC distributions of the experimental data are
seen to be broader than both the mixed and V+G events. The result
again suggests the presence of non-statistical fluctuations.

\begin{figure}
\vspace*{-0.4cm}
\epsfxsize=8cm
\centerline{\epsfbox{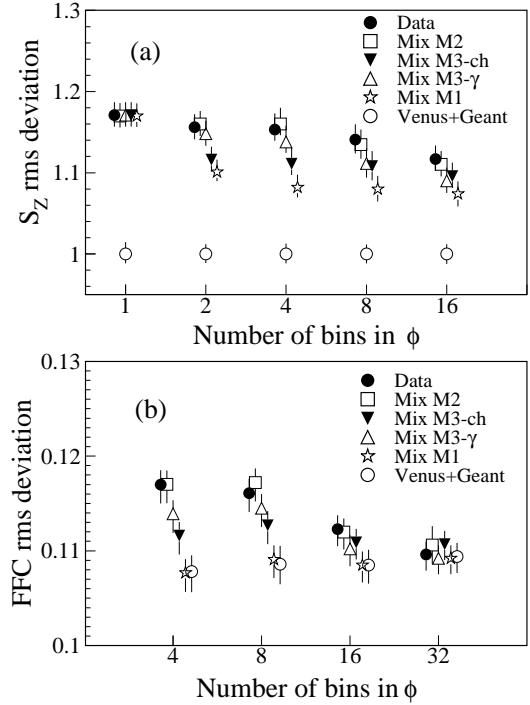}}
\vspace*{-1cm}
\caption {\label{sz_ffc_rms}
The root mean square (rms) deviations of the $S_Z$  and FFC distributions
for various divisions in the azimuthal angle.
}
\end{figure}

   The rms deviations of the $S_Z$ and FFC distributions as a function of the 
   number of bins in azimuth is shown for experimental data, mixed events, and
   V+G in  Fig.~\ref{sz_ffc_rms}.
   The statistical errors on the values are small and lie within
   the size of the symbols.
   The error bars include both statistical and systematic errors.
   The systematic errors include effects such as uncertainty in the
   detection efficiencies, gain fluctuations, backgrounds, binning
   variations and fitting procedures. The total systematic error was
   obtained as the sum in quadrature of the individual error contributions.

   Since the mixed events are constructed to maintain
   the $N_{\gamma-{\mathrm like}}$--$N_{\mathrm ch}$ 
   correlations for the full azimuth (bin~1),
   the rms deviations of data and mixed events for this bin
   are identical. The difference of the rms deviations 
   between data and V+G for this bin is the same as shown in
   the right hand panel of Fig.~\ref{dcc_global} \cite{WA98-3}.
   The effect of correlations due to charged particle contamination in
   $N_{\gamma-{\mathrm like}}$ sample mentioned above 
   (see Fig.~\ref{m1_m2_check})
   has been removed by rescaling the mixed event results according to the 
   percentage difference between the rms deviations of the $S_Z$ and
   FFC distributions of V+G events and those 
   of the corresponding V+G mixed events.

Comparison of the rms deviations of the experimental data and mixed events
have been made for all the three different cases.
\begin{itemize}

\item (1) {\bf Comparison with M1} : From Fig.~\ref{sz_ffc_rms}
one notices that
the data points are higher compared to those of the M1 events for
several bins in $\phi$.
The differences between the data points and M1 mixed events have been
quantified by using the quantity, 
$\sigma=\sqrt{ (\sigma_{lower}^{data})^2 + (\sigma_{upper}^{M1})^2 }$.
For 2, 4, and 8bins the values of the $S_Z$ rms deviations of the data are 
2.5$\sigma$, 3.0$\sigma$, 2.4$\sigma$ larger compared to those of the
M1 events, respectively. Similarly the rms deviations of the FFC distributions
at 4 and 8 bins for data are 3.7$\sigma$ and 2.8$\sigma$ larger than
those of the M1 events. At 16 and 32 bins the results for mixed events
and data agree within the quoted errors.

This indicates the presence of localized non-statistical fluctuations.
This result is completely model independent as the comparison has been 
made to mixed events generated from data.
The observed non-statistical fluctuations may arise because of (a)
event-by-event correlated $N_{\gamma-{\mathrm like}}$-$N_{\mathrm ch}$ 
fluctuations, (b) individual fluctuations in 
$N_{\gamma-{\mathrm like}}$ and (c) individual fluctuations $N_{\mathrm ch}$.
The source of the fluctuation can be obtained by comparing the data
with the results from M2, M3-$\gamma$ and M3-ch.

\item (2) {\bf Comparison with M2} : From Fig.~\ref{sz_ffc_rms} it is
evident that the rms deviations of the M2
events agree with those of the experimental data within errors for all 
bins in $S_Z$ and FFC distributions. This implies the absence of event-by-event
correlated fluctuations in $N_{\gamma-{\mathrm like}}$ versus $N_{\mathrm ch}$.

If the amount of DCC-like fluctuations in the experimental data were large,
then the rms deviations for data would have been larger than those of M2
events. Since this is not the case, we have extracted upper limits on
the probability of DCC-like fluctuations at the 90\% confidence level
by comparing the results from data with those obtained from the 
nDCC events. Using the DCC-model used in this analysis we have
extracted upper limits on the probability of DCC events to be
10$^{-2}$ for $\Delta\phi$ between
45-90$^\circ$ and 3$\times$10$^{-3}$ for $\Delta\phi$ between 90-135$^\circ$.

\item (3) {\bf Comparison with M3-$\gamma$ and M3-ch} : The M3 type
mixed events shown in Fig.~\ref{sz_ffc_rms} are found to be similar 
to each other within the quoted errors and lie between M1 and M2. 
This indicates the presence of localized independent fluctuations 
in $N_{\gamma-{\mathrm like}}$ and $N_{\mathrm ch}$. 

\end{itemize}

\section{Summary}

Multiplicity distributions and fluctuations are important for
understanding the evolution of nuclear systems at high energy
collisions. These have been studied from the multiplicity
measurements of charged particles and photons in
158$\cdot A$ GeV Pb+Pb reactions. 
Production of charged particles and photons over the
full range of centrality could be described in terms of participant
scaling as $N_{\rm part}^{1.07\pm 0.05}$ and 
$N_{\rm part}^{1.12\pm 0.03}$, respectively. 
This indicates violation of the naive wounded nucleon model.
Multiplicity fluctuations have
been studied by varying the centrality of the reaction. It has been
shown that the centrality of the reaction has to be properly
chosen so that the multiplicity distributions are of
good Gaussians. We have shown that this criterion is satisfied by
making narrower bins in $E_{\rm T}$ corresponding to 2\% of minimum bias 
cross sections. The fluctuations for photons are constant 
($\omega=2.0\pm 0.06$) over the centrality range considered and
are higher than those for charged particles ($\omega=1.8\pm 0.11$).

Event-by-event charged to neutral fluctuations have been studied
in full acceptance of the detectors and in localized ($\eta-\phi$)
phase space. Using global event characterization,
no event with large charged-neutral fluctuations have been observed.
A mixed event analysis is not possible for this global search.
The search for localized fluctuations have been carried out
by comparing data with mixed events generated from the data.
Full understanding of the nature of the mixed events have been
achieved by using a simple DCC model.
Two different analysis methods, (1) 
$N_{\gamma-{\rm like}}$ and $N_{\rm ch}$ correlation method, and (2) a more
sophisticated DWT method, have been employed. Both analysis methods
provide model-independent evidence for non-statistical fluctuations
at the 3$\sigma$ level for $\phi$ intervals greater than 45$^\circ$.
This is seen to be due to non-statistical fluctuations in both
both $N_{\gamma}$ and $N_{\rm ch}$.
No significant event-by-event correlated fluctuations are observed,
contrary to expectations for a DCC effect. The origin of the 
observed individual fluctuations are not known at present.
The interpretation of the results remains an open question.

With much higher particle multiplicities achieved at higher energies 
of RHIC and LHC, analysis 
methods using event-by-event fluctuations would certainly 
be very essential to probe the signals of QGP and DCC. 
Better understanding 
of data at SPS energies would definitely help in making firmer conclusions of 
any signal from future experiments.


\begin{thebibliography}{99}
\bibitem{qm99}      Proceedings of {\it Quark Matter '99},
                    Nucl. Phys A {\bf 661} (1999).
\bibitem{qm01}      Proceedings of {\it Quark Matter 2001},
                    Stoney Brook, USA, to be published in Nucl. Phys. A.
\bibitem{shuryak} E.V. Shuryak and O. Zhirov, Phys. Lett. {\bf B89} (1980) 253.
\bibitem{vanhove} L. van Hove, Phys. Lett. {\bf B118} (1982) 138.
\bibitem{step}    M. Stephanov, et al., Phys. Rev. 
                       Lett. {\bf 81}, 4816 (1998).
\bibitem{heisel}   Gordon Baym and Henning Heiselberg,
                    Phys. Lett. {\bf B469}, 7 (1999).
\bibitem{asakawa}      M.Asakawa, U.Heinz, and B.M{\"u}ller, Phys. Rev. Lett.
                    {\bf 85}, 2072 (2000).
\bibitem{jeon}      S. Jeon and V. Koch, Phys. Rev. Lett.
                    {\bf 85}, 2076 (2000).
\bibitem{anselm}    A.A. Anselm, M.G. Ryskin, 
                      Phys. Lett. B {\bf 266}, 482 (1991).
\bibitem{bj}        J.D. Bjorken, Int. J. Mod. Phys. A {\bf 7}, 4189 (1992).
                      J.D. Bjorken, K.L. Kowalski, C.C. Taylor, 
                      ``Baked Alaska'', SLAC-PUB-6109, Apr. 1993.
\bibitem{blaizot}   J. -P. Blaizot and A. Krzywicki, 
                      Phys. Rev. D {\bf 46}, 246 (1992).
\bibitem{raj}       K. Rajagopal and F. Wilczek, 
                      Nucl. Phys. B {\bf 399}, 395 (1993); 
                      Nucl. Phys. B {\bf 404}, 577    (1993).
\bibitem{huang}     Z. Huang, et al.,
                       Phys. Rev. D {\bf 54}, 750 (1996).
\bibitem{qm97nayak} T.K. Nayak, WA98 Collaboration, 
                    Nucl. Phys. {\bf A638}, 249c (1998).
\bibitem{dccflow}   B.K.~Nandi, G.C.~Mishra, B.~Mohanty, D.P.~Mahapatra
                       and T.K.~Nayak, Phys. Lett. {\bf B449} (1999) 109.
\bibitem{dccstr}    B.K.~Nandi, T.K.~Nayak, B.~Mohanty, D.P.~Mahapatra
                       and Y.P.~Viyogi, Phys. Lett. {\bf B461} (1999) 142.
\bibitem{randrup}  J. Randrup, Phys. Rev. C {\bf 62}, 064905 (2000). 
\bibitem{WA98-9}    WA98 Collaboration, M.M.~Aggarwal et al., 
                       Phys. Lett. B {\bf 458}, 422 (1999).
\bibitem{wa98nim}   M.M.~Aggarwal et al., 
                       Nucl. Instr. and Meth. A {\bf 424}, 395 (1999).
\bibitem{venus}     K. Werner, Phys. Rep. {\bf 232}, 87 (1993).
\bibitem{bialas}     A. Bialas, A. Bleszynski and W. Czyz, Nucl. Phys. B 111, 
                     2000 (1976).
\bibitem{WA98-10}    WA98 Collaboration, M.M.~Aggarwal et al., 
                     Eur. Phys. J {\bf C18}, 651 (2001).
\bibitem{WA98-3}    WA98 Collaboration, M.M.~Aggarwal et al., 
                    Phys. Lett. B {\bf 420}, 169 (1998).
\bibitem{WA98-12} WA98 Collaboration, M.M.~Aggarwal et al., 
                  e-print archive: nucl-ex/0012004.
\bibitem{amara}     A Graps, IEEE Computational Sciences and Engineering, 
                       {\bf 2}, 50 (1995).
\bibitem{numeric}   Numerical Recipes, Cambridge Univ. Press , 1998.
\end{thebibliography}
\end{document}